\documentstyle{article}
\title{Current-Induced Chiral Phase Transition in a Josephson-Junction Array}
\author{M.V. Simkin \\ {\em Department of Physics, Brown University,}\\
{\em Providence, RI 02912-1843}}
\date{ }
\begin{document}
\maketitle
\begin{abstract}
It is predicted that the fully frustrated Josephson-junction array on a 
square lattice
with alternating  columns of junctions having a reduced critical current
$I_{weak}=\mu I_{0}$(with  $\mu<1$ and  $I_{0}$ the critical current of
all other junctions) in certain range of $\mu$ will under increasing
of the external current first change from the state with zero chirality
to the state with nonzero one at the lower critical current $I_{c1}$ before
entering  nonsuperconducting state at the upper critical current $I_{c2}$.
\end{abstract}

A square two-dimensional periodic Josephson-junction array (JJA) in a uniform
transverse  magnetic field, with half a flux
quantum per plaquette, is a realization of a fully frustrated (FF) XY model 
\cite{vil}.
Its ground state is a checkerboard pattern of plaquettes with currents
flowing clockwise and anti-clockwise \cite{tei}. The chirality, defined 
as the sum
of currents around plaquette in a clockwise direction, has antiferromagnetic
order.
An interesting  generalization of the FF XY model was proposed by Berge 
{\it et al}\cite{ber}. For the case of JJA it corresponds to an array
in which the critical currents of alternate columns are reduced 
$I_{weak}=\mu I_{0}$,where $I_{0}$ is the critical current of
all other junctions, and $\mu<1$. See Fig.1.
For $\mu<1/3$ the ground state has zero chirality,
while for $\mu>1/3$ it has a nonzero chirality.
In this article  I investigate the transformation of the ground state under
an external current applied in the direction transverse to the weak bonds.

When the zero current ground state for $\mu<1/3$ has zero chirality 
one expects  that when we apply the external current the chirality will 
remain zero. 
The ground state with current will look like Fig.1 with all $\theta=0$, except
there will be a phase difference $\phi$ at each horizontal junction to
accommodate current $I=I_0\sin\phi$. However, for big enough current this
state may become unstable against fluctuations leading to nonzero chirality.
To check this we write down the energy per plaquette as a function of 
deviations $\theta_i$ from the ground state values
of phases of the  superconducting nodes \cite{gauge}:
\begin{equation}
E=-\cos(\phi + \theta_1 -\theta_2) - \cos(\phi + \theta_3 -\theta_4)-
\cos(\theta_1 -\theta_3) + \mu \cos(\theta_2 -\theta_4).
\end{equation}
After straightforward but tedious algebra \cite{algebra} we find that the  
nonchiral state is stable for currents $I<I_{c1}$, given by 
\begin{equation}
I_{c1}=\frac{1}{1-\mu}\sqrt{1-2\mu-3\mu^2}.
\end{equation}
As we expect, $I_{c1}(\mu=0)=1$, as we than effectively have an unfrustrated 
array, and $I_{c1}(\mu=1/3)=0$, as for $\mu>1/3$ even the zero current ground 
state is chiral.

We now determine the upper critical current, i.e. the maximum current the
array can sustain with zero voltage, but allowing for nonzero chirality,
following the method of Benz {\it et al}\cite{bnz} for a 
uniform ($\mu=1$) array.

A chiral current-carrying state is shown in Fig.2. The phase differences
across the junctions $\alpha,\beta,\gamma,\delta$ obey the following
conditions. 
The sum of currents flowing into a node is zero:
\begin{equation}
\sin\alpha-\sin\gamma-2\sin\beta=0.
\end{equation}
Net current in vertical direction is zero:
\begin{equation}
\sin\beta=-\mu\sin\delta.
\end{equation}
The sum of phase differences around plaquette is zero:
\begin{equation}
\alpha-\gamma+\beta+\delta=0.
\end{equation}

The net current per junction in the horizontal direction is 
\begin{equation}
I=\frac{1}{2}(\sin\alpha+\sin\gamma).
\end{equation}
From (3)-(6) we obtain:
\begin{equation}
I^2=\frac{a^2-2ax}{a^2-2ax+x}x,
\end{equation}
where $a=\frac{1}{2}(\frac{1}{\mu}+1)$, $x=\cos^2(\frac{\alpha-\gamma}{2})$.
We now maximize $I$, given by (7), with the restriction $x \le 1$.
Straightforward, though tedious, analysis shows that for 
$\mu > \mu_c = 0.2137...$ \cite{com} the maximum current $I_{c2}$ is achieved 
for some value of $x<1$ and is given by
\begin{equation}
I_{c2}=\frac{1}{2}\frac{\frac{1}{\mu}+1}{\sqrt{\frac{1}{\mu}+1}+1}.
\end{equation}
For $\mu=1$ Eq. (8) gives the result of Benz {\it et al}\cite{bnz} 
$I_{c2}=\sqrt{2}-1$.
For $\mu < \mu_c$,the  maximum $I$ is achieved for $x=1$, or $\alpha=\gamma$
which is  a nonchiral state. Correspondingly for $\mu < \mu_c$,
$I_{c2}=I_{c1}$.

In Fig.3 the phase diagram of the nonuniform FF JJA in the ($\mu,I$) plane
is shown. We see that for $0.2137... =\mu_c<\mu<1/3$ the array as the current
is increased will first change from a nonchiral 
to a chiral state at $I=I_{c1}$, and then to a non-superconducting state
at $I=I_{c2}>I_{c1}$.

From Eqs. (2) and (7) we obtain for current $I$ just above $I_{c1}$ the
chirality 
\begin{equation}
\chi=\sin\alpha-\sin\gamma+\sin\beta-\mu\sin\delta,
\end{equation}
behaves as
\begin{equation}
\chi \sim (I-I_{c1})^\frac{1}{2}.
\end{equation}

I observed the above transition in computer simulations (see Fig.4).
For the simulations I used, following Falo {\it et al} \cite{fal},a 
resistively 
shunted Josephson-junction array,
with each superconducting node having a capacitance to the ground.
The current was injected into  16$\times$16 array from superconducting bars 
at the 
edges. Free boundary conditions where used in the transverse direction.
Very small thermal fluctuations ($10^{-8}$ in units of Josephson energy 
$\hbar I_0/2e$) are introduced to drive the system away from the unstable
nonchiral state and  to check the chiral state for stability.
In Fig. 4(a),(b) the phases of the superconducting nodes for the different
applied currents are represented by arrows.
In Fig. 4(a) $\mu=0.32<1/3$,
$I=0.3<I_{c1}$, and the phases in one column are identical which is expected
for zero chirality. In Fig. 4(b) the current $I=0.5>I_{c1}$, the phases in a
column are varying which indicates a finite chirality, which is shown
in  Fig. 4(c). 

~
In conclusion I have obtained the phase diagram of nonuniform 
fully-frustrated Josephson array in the applied current($I$)-weak bond
 strength($\mu$) plane.
For some range of $\mu$ it implies a transition from nonchiral to chiral
state under increasing of the applied current. I have observed this
transition in numerical simulations. 

I am grateful to N.Akino, J.M. Kosterlitz, and J.B. Marston for useful 
conversations. This work was supported by National Science Foundation
Grant No. DMR-9222812.

\newpage
\begin {figure}
\caption{Nonuniform fully-frustrated Josephson-junction array in a nonchiral 
current-carrying state. Solid lines are junctions with
strong critical current, $I_{0}$. Dashed lines - those with weak critical
current $\mu I_{0}$. Arrows - directions of currents. The phases of the 
superconducting nodes are indicated. In a nonchiral state all $\theta=0$,
though this state may be unstable against fluctuations of $\theta$. }
\end {figure}
\begin {figure}
\caption{Nonuniform fully-frustrated Josephson-junction array in a chiral
state. Thick lines with arrows - directions of currents. 
$\alpha,\beta,\gamma,$
and $\delta$ - phase differences across junctions.}
\end {figure}
\begin {figure}
\caption{Phase diagram of the nonuniform fully-frustrated Josephson-junction 
array in the ($\mu,I$) plane, computed using Eqs.(2) and (8).}
\end {figure}
\begin {figure}
\caption{Results of numerical simulations for a $16\times 16$ array with the
weak bonds strength $\mu=0.32$. In (a) and (b) phases of superconducting
nodes are represented by arrows. (a) corresponds to current $I=0.3$ and
is in a nonchiral part of the diagram (see Fig.3). (b) $I=0.5$, which
is bigger than the critical current for the chiral transition.
(c) is (b) in chirality (sum of currents around the plaquette) representation.}
\end {figure}

\begin{thebibliography}{99}
\bibitem{algebra} I just expand $E$ in $\theta$ to the second order
and check whether the resulting quadratic form is positive definite.
\end{thebibliography}
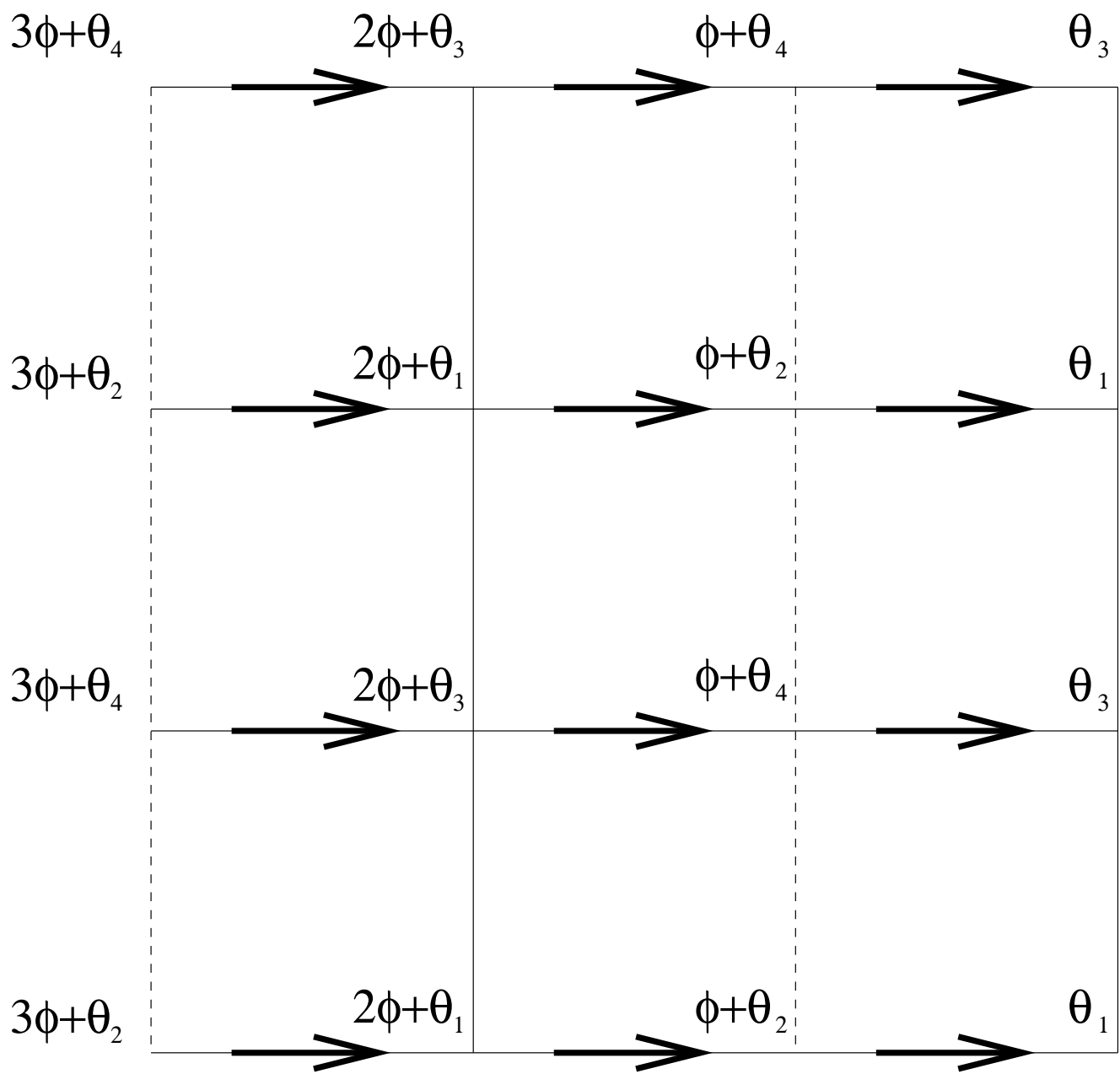
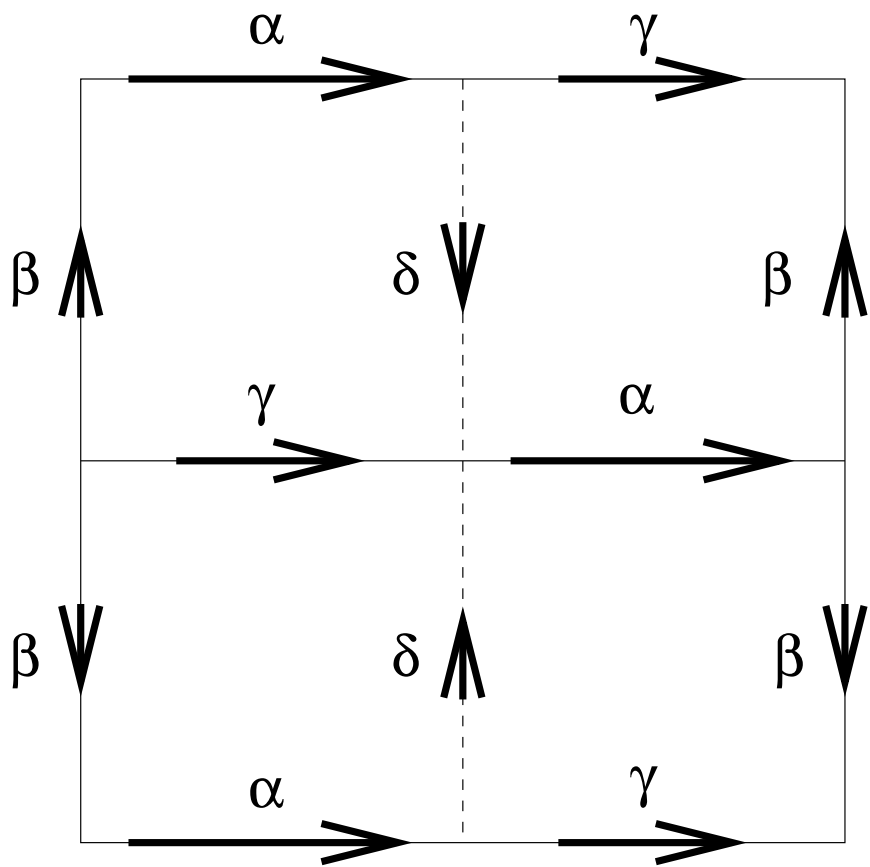
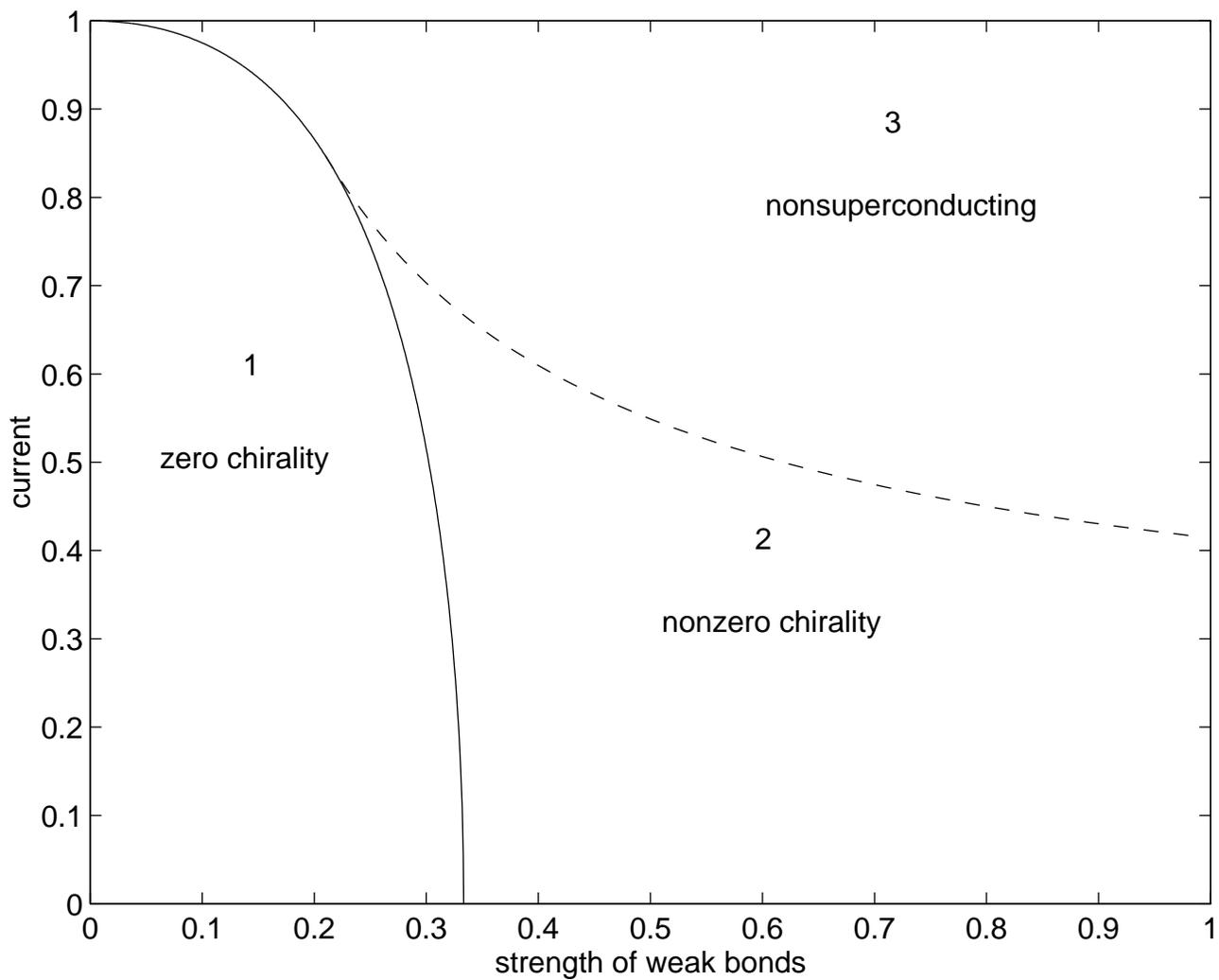
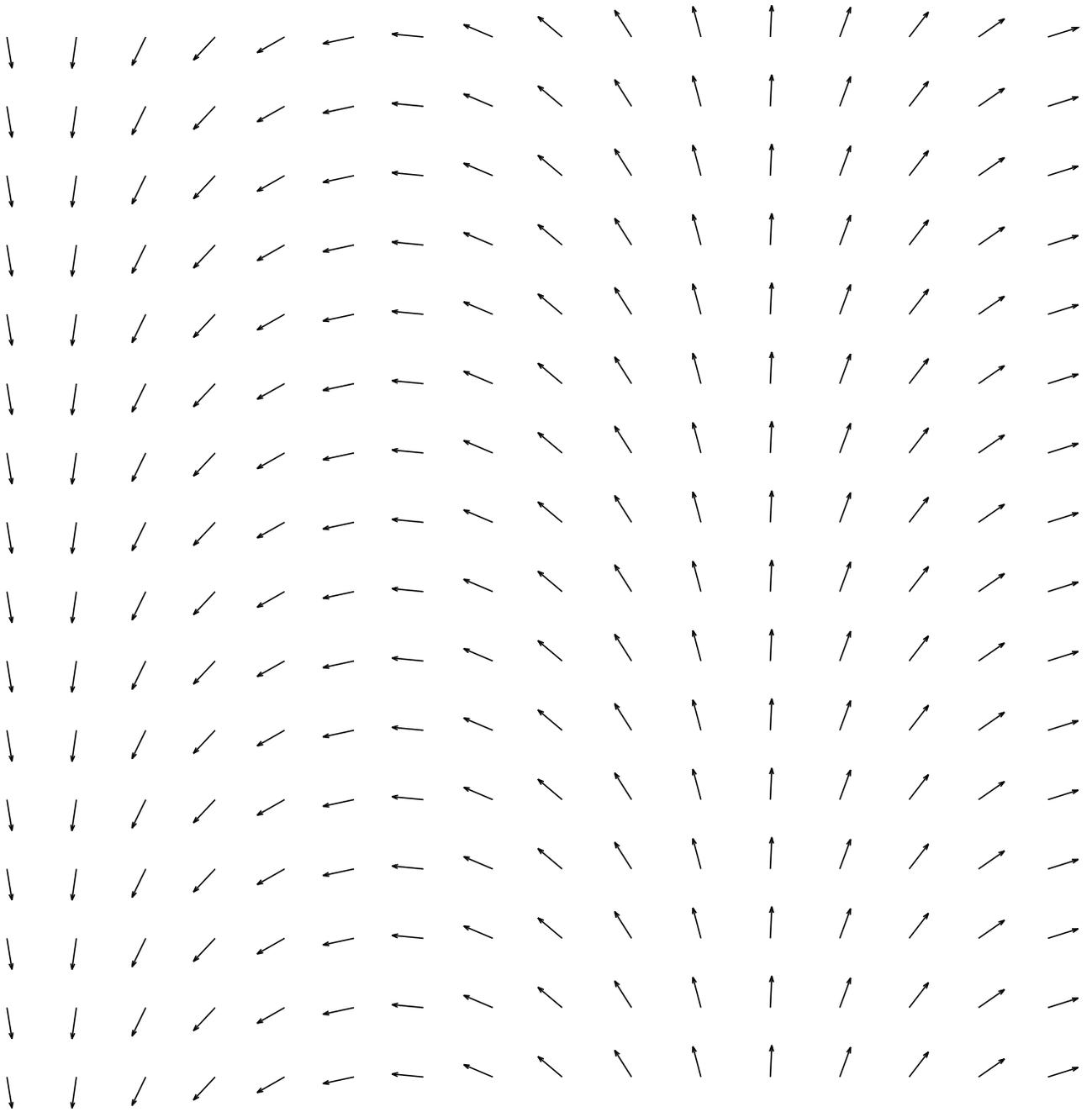
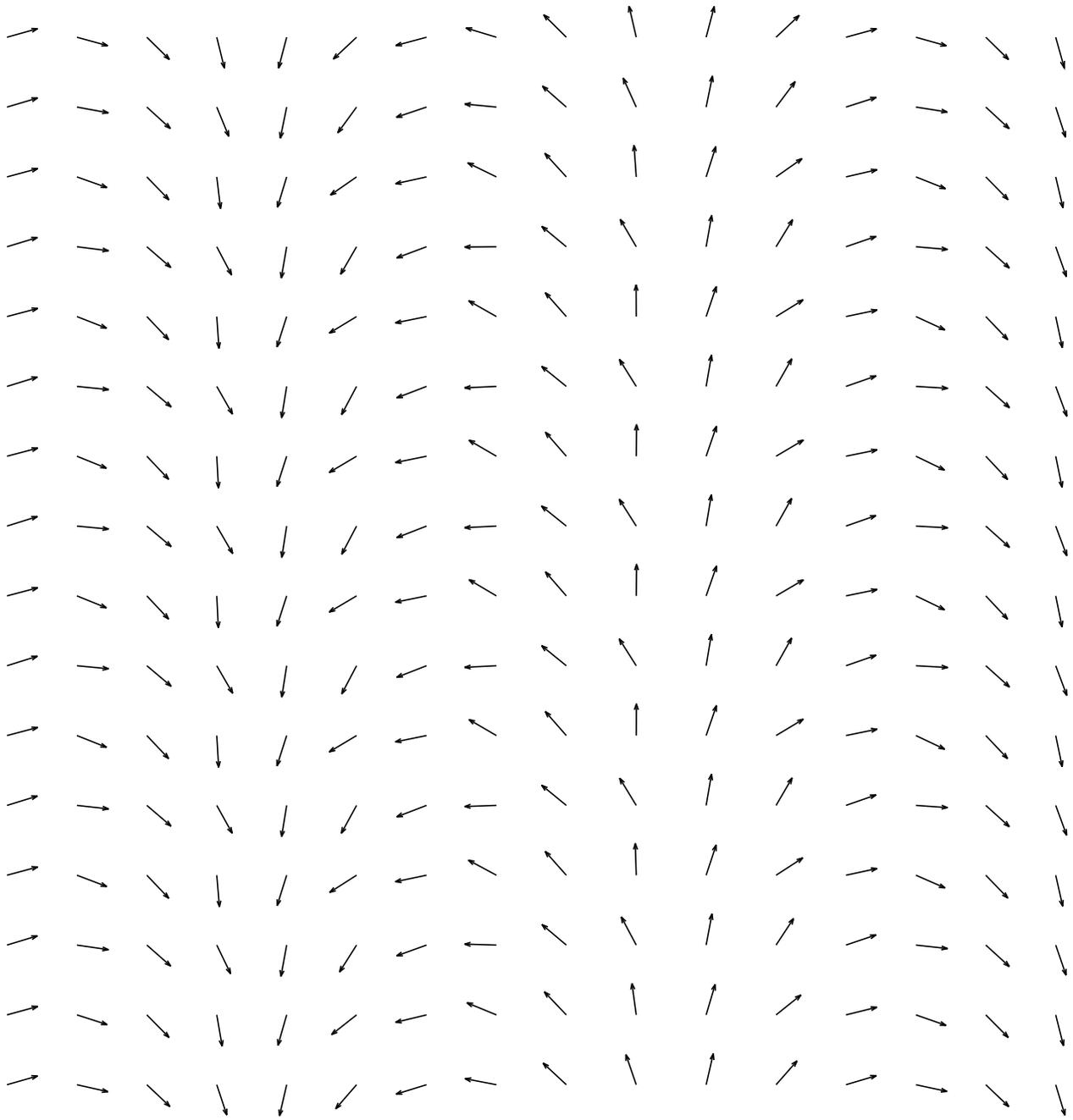
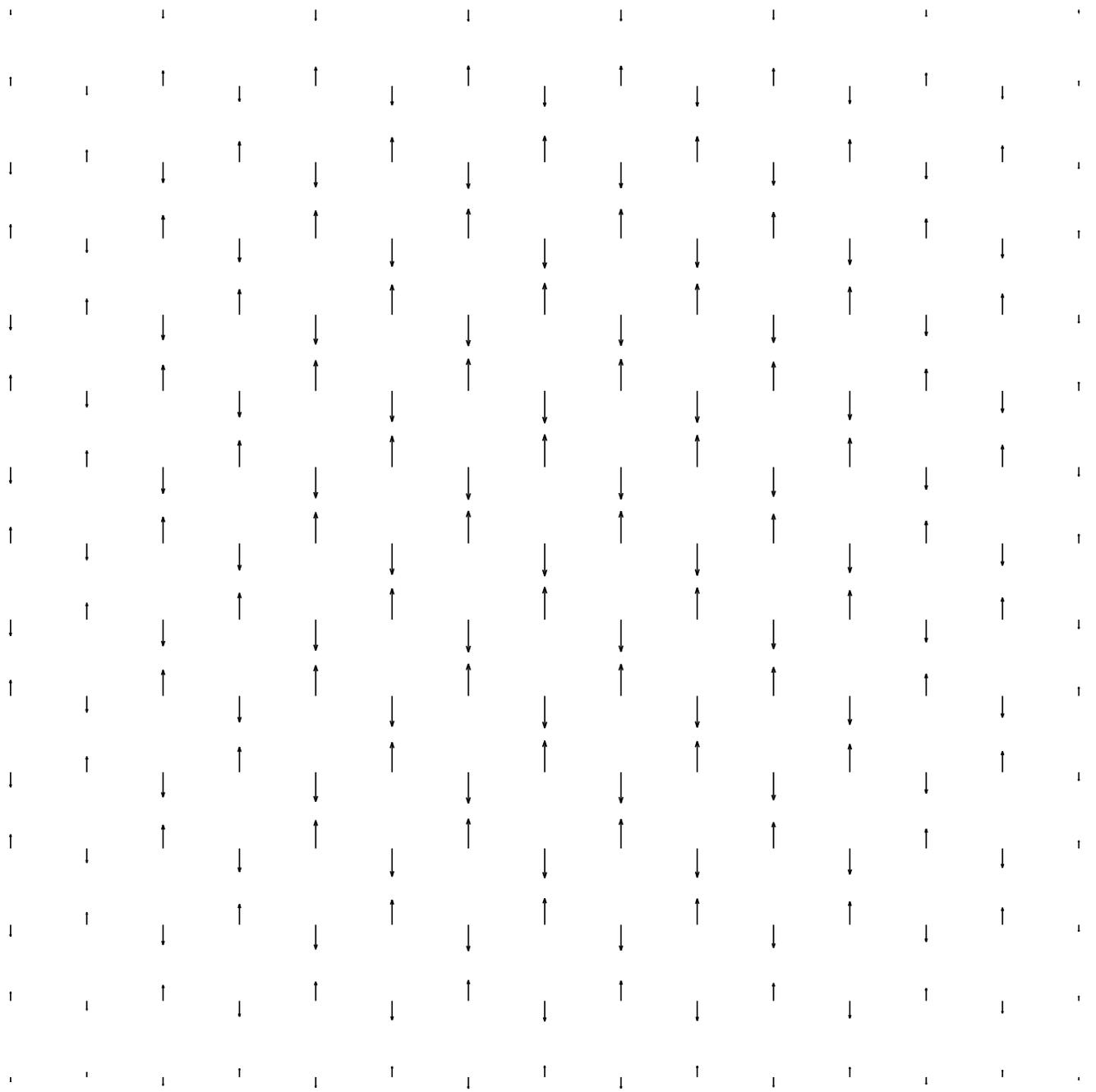
\end{document}